\documentclass[twocolumn,pra,amsmath,amssymb,unsortedaddress,showpacs,floatfix,citeautoscript,nofootinbib]{revtex4-1}

\usepackage{latexsym}
\usepackage{dcolumn}
\usepackage{graphicx}
\usepackage{subfigure}
\usepackage{color}
\usepackage[colorlinks,urlcolor=blue,citecolor=blue]{hyperref}

\begin{document}

\title{Multiple signatures of topological transitions for interacting fermions in chain lattices}
\author{Y.-H. Chan}
\affiliation {Institute of Atomic and Molecular Sciences, Academia
Sinica, Taipei 10617, Taiwan}
\author{Ching-Kai Chiu}
\affiliation {Department of Physics and Astronomy, University of
British Columbia, Vancouver, British Columbia, Canada V6T 1Z1}
\author{Kuei Sun}
\thanks{Corresponding author. kuei.sun@utdallas.edu}
\affiliation {Department of Physics, The University of Texas at
Dallas, Richardson, Texas 75080-3021, USA}


\begin{abstract}
We study one-component fermions in chain lattices with
proximity-induced superconducting gap and interparticle
short-range interaction, capable of hosting Majorana fermions. By
systematically tracking various physical quantities, we show that
topological states and topological phase transitions in the system
can be identified by multiple signatures in thermodynamic
quantities and pair-condensate properties, in good agreement with
the known signatures in the ground-state energy and entanglement
spectrum. We find the disappearance of the topological phase in a
largely attractive regime, in which the system undergoes a
first-order transition between two topologically trivial states.
In addition, the stability of the signatures against finite size,
disorder, and inhomogeneity is analyzed. Our results provide
additional degrees of freedom for the characterization of
topological states with interaction and for the experimental
detection of emergent Majorana fermions.
\end{abstract}

\maketitle

\vspace{-0.3cm}
\section{Introduction}\label{sec:introduction}
\vspace{-0.3cm}

Exploring topological states of matter has become a rapidly
developing field in condensed matter
physics~\cite{Hasan10,Qi11,Chiu15}. One intriguing topological
state that exhibits both fundamental interest and practical
application is the emergent Majorana zero modes or Majorana
fermions~\cite{Wilczek09,Franz10,Alicea12,Beenakker13,Elliott15},
which are their own antiparticles and possess zero energy, in
superconducting materials. The pursuit of the Majorana modes
started from the study of a one-dimensional (1D) $p$-wave
superconducting chain by Kitaev~\cite{Kitaev01}. This milestone
has triggered several alternative schemes for the realization of
Majorana fermions, such as $p+ip$
superconductors~\cite{Stone04,Fendley07,Tewari07,Raghu10,Chung11},
spin-orbit coupled superconducting
nanowires~\cite{Lutchyn10,Oreg10,Cook11,Stanescu13}, chains of
magnetic atoms on superconducting
substrates~\cite{Nadj-Perge13,Nadj-Perge14,Hui14,Dumitrescu15},
superconducting surfaces of topological
insulators~\cite{Fu08,Linder10,Chiu12,Hung13}, superfluid helium
3~\cite{Kopnin91,Qi09,Chung09}, and ultracold
atoms~\cite{Liu12b,Qu13,Zhang13,Chen13,Ruhman15,Jiang15}.
Recently, a semiconductor nanowire with intrinsic spin-orbit
coupling, external magnetic field, and proximity-induced
superconductivity has become one of the experimentally promising
platforms to host Majorana
fermions~\cite{Mourik12,Deng12,Rokhinson12,Das12,Finck13,Churchill13},
which appear on the edges of the wire and lead to a tunneling
conductance peak at zero voltage. Such a zero-bias peak has been
observed as (indirect) evidence for their existence.

In addition to the transport
properties~\cite{Bolech07,Law09,Qu11,Liu12,Lin12,Roy12,Jacquod13,Liu13,Li14,Wu14b,Doornenbal15,Setiawan15,Lopes15},
searching different signatures for the topological phase is an
ongoing task for the investigation of Majorana fermions. From the
experimental point of view, it not only provides more evidence for
direct detection of Majorana fermions but also helps rule out
different physical causes that result in the same transport
behavior~\cite{Lee12,Kells12,Sun15,Bercioux15}. From the
theoretical point of view, a comparison between various
quantities, even those typically used to describe
Ginzburg-Landau-type phase transitions, such as susceptibilities
and superfluid order, can provide useful information for
characterizing topological phases and topological phase
transitions, especially in interacting
systems~\cite{Rombouts10,Sun14a,Amaricci15}. Interaction effects,
which are unavoidably present in reality, may alter physical
features of the topological phase and even change the
topology~\cite{Rombouts10,Sun14a,Amaricci15,Stoudenmire11,Thomale13,Lin13,Ortiz14,Chan15,Keselman15,Sela11,Hassler12,Hung13b,Fidkowski10a}.
For example, time-reversal symmetric Kitaev chains in class BDI
change the topological invariant from $\mathbb{Z}$ to
$\mathbb{Z}_8$ when the interaction is turned
on~\cite{Fidkowski10a}. In an interacting system, Majorana zero
modes at edges become many-body Majorana wave
functions~\cite{Goldstein12,Kells14} and the degenerate ground
states with two different parities are connected by these
many-body Majorana zero operators. Such many-body phenomena
attract broad interest, from the fundamental understanding of its
nature to applications on quantum
computation~\cite{Kitaev03,Nayak08,Alicea11,Laflamme14}. The study
of multiple signatures shall provide a convincing series of tests
to characterize the interacting topological phase diagram. Recent
works have analyzed individual quantities for separate models,
such as entanglement spectrum in a spin-orbit coupled chain with
interactions~\cite{Stoudenmire11,Chan15},
compressibility~\cite{Nozadze15} and spectral
function~\cite{Thomale13} in the Kitaev chain, as well as
susceptibility~\cite{Ortiz14} and pair
correlation~\cite{Rombouts10} in long-range coupled
superconducting fermions. However, these works have not provided a
systematic comparison of multiple quantities between
topological/trivial phases or upon topological transitions within
a single frame (model).

In this paper, we study various physical quantities of 1D
one-component fermions having proximity-induced pairing gap and
interparticle short-range interaction, as a generalization of the
Kitaev model. These quantities are obtained from
density-matrix-renormalization-group
(DMRG)~\cite{White92,Schollwock05} calculations and categorized
into three groups: (i) topological properties, including
ground-state degeneracy and entanglement spectrum; (ii)
thermodynamic properties, including compressibility and
susceptibility; and (iii) condensate properties, including
pair-condensate fraction and Cooper-pair size. We take the
topological region indicated by the first group as a reference and
investigate the behavior of the second and third ones in the
parameter space. As a result, we shall find alternative signatures
to identify the topological states and topological transitions. In
addition, by tracking the multiple signatures, we show the
topological phase diagram as a function of interaction. Finally,
we study the behavior of several signatures against finite size,
disorder, as well as inhomogeneity, and discuss their stability
under these conditions.

The paper is organized as follows. In Sec.~\ref{sec:model}, we
introduce the model Hamiltonian and define the physical quantities
of interest. Section \ref{sec:numerics} shows the setup of DMRG
calculations. We present the results and discussions in
Sec.~\ref{sec:results}. Section \ref{sec:conclusion} is the
conclusion.

\vspace{-0.3cm}
\section{Model and physical quantities}\label{sec:model}
\vspace{-0.3cm}

In this section, we present the model in consideration and
physical quantities of interest. We consider one-component
fermions in chain lattices where particles can scatter through the
Cooper channel (or form Cooper pairs) due to a combined effect of
external proximity-induced pairing and internal short-range
interaction. If there is only the external effect, the Hamiltonian
is of the Kitaev form~\cite{Kitaev01},
\begin{eqnarray}
{H_{{\rm{K}}}} = {H_0} + \sum\limits_{j = 1}^{L - 1} {\left(
{\Delta ' \hat c_{j + 1}^\dag \hat c_j^\dag  +
{\rm{H}}{\rm{.c}}{\rm{.}}} \right)}, \label{eq:Kitaev}
\end{eqnarray}
where ${\hat c_j^\dag }$ creates a fermion on site $j$, $\Delta '$
(taken real without the loss of generality) describes the
proximity-induced pairing, $L$ is the total number of lattice
sites, and
\begin{eqnarray}
{H_0} = \sum\limits_{j = 1}^{L - 1} { - t\left( {\hat c_j^\dag
{{\hat c}_{j + 1}} + {\rm{H.c.}}} \right)}  - \mu \sum\limits_{j =
1}^L {{{\hat n}_j}}\label{eq:free_H}
\end{eqnarray}
is the noninteracting Hamiltonian with number operator ${{\hat
n}_j} = \hat c_j^\dag {{\hat c}_j}$, nearest-neighbor tunneling
$t$, and chemical potential $\mu$. If there is only the internal
interaction effect, the Hamiltonian reads
\begin{eqnarray}
{H_{\rm{I}}} = {H_0} + \sum\limits_{j = 1}^{L - 1} {V'{{\hat
n}_j}{{\hat n}_{j + 1}}},
\end{eqnarray}
where negative (positive) $V'$ represents attractive (repulsive)
interaction. In realistic cases of a
nanowire~\cite{Mourik12,Deng12,Rokhinson12,Das12,Finck13,Churchill13}
or an atomic chain~\cite{Nadj-Perge13,Nadj-Perge14,Hui14} on a
superconducting substrate, or a tube of quantum gases in
higher-dimensional optical lattices~\cite{Jiang15}, both effects
can take place. We hence model the system with a phenomenological
parameter $\gamma$ to describe the relative strength of the two
effects and write down our model Hamiltonian,
\begin{eqnarray}
H &=& (1 - \gamma ){H_{\rm{K}}} + \gamma {H_{\rm{I}}}\nonumber\\
&=&{H_0} + \sum\limits_{j = 1}^{L - 1} {\Delta \left( {\hat c_{j +
1}^\dag \hat c_j^\dag  + {\rm{H}}{\rm{.c}}{\rm{.}}} \right) +
V{{\hat n}_j}{{\hat n}_{j + 1}}}, \label{eq:Hamiltonian}
\end{eqnarray}
with independent variables $\Delta  = (1 - \gamma )\Delta '$ and
$V=\gamma V'$, which can be self-consistently determined by
microscopic degrees of freedom and/or realistic parameters of the
system. (Below we aim to study physical characteristics of $H$ in
a range of its parameter space rather than determine the
parameters for a specific situation.) The Hamiltonian always
conserves the even/odd parity of total number of particles $N =
\langle {\hat N} \rangle = \langle {\sum {{{\hat n}_j}} } \rangle
$ due to $[ {H,{{( - 1)}^{\hat N}}} ] = 0$ and conserves $N$
(meaning $[ {H,\hat N} ] = 0$) in the limit $\Delta \to 0$.

We will study the behavior of three groups of physical quantities
of interest in the topological and non-topological regions as well
as upon the topological transition. The first group, called
topological quantities, includes ground-state degeneracy and
entanglement spectrum degeneracy. The former one can be
characterized by the energy difference
\begin{eqnarray}
\delta E = |E_{\rm {even}}-E_{\rm {odd}}| \label{eq:delta_E}
\end{eqnarray}
between the lowest eigen energies ${E_{{\rm{even/odd}}}}$ of the
even and odd blocks of the Hamiltonian [where ${\langle {{{( -
1)}^{\hat N}}} \rangle _{{\rm{even/odd}}}} = \pm 1$],
respectively. The ground-state degeneracy $\delta E =0$ occurs as
the manifestation of Majorana fermions in the topological region
and does not otherwise.

The entanglement spectrum is a series of eigenvalues of reduced
density matrix $\rho_{\rm{R}}$ obtained by tracing out half
spatial degrees of freedom of the ground-state wave function $|
{{\psi _{\rm{g}}}} \rangle $,
\begin{eqnarray}
{\rho _{\rm{R}}} = {\rm{T}}{{\rm{r}}_{j \le L/2}}\left| {{\psi
_{\rm{g}}}} \right\rangle \left\langle {{\psi _{\rm{g}}}} \right|,
\end{eqnarray}
where we consider $L$ an even number and choose the ground state
of the even-particle-number block when the even-odd degeneracy
occurs ($\delta E =0$). We obtain a set of $2N_{\lambda}$ largest
eigenvalues of ${\rho _{\rm{R}}}$, denoted by $\{ \lambda_j \}$
and sorted as $\lambda_j \ge \lambda_k$ if $j<k$, and compute the
difference between paired elements defined by
\begin{eqnarray}
\delta \lambda  \equiv \sum\limits_{j=1}^{N_{\lambda}} {\left|
{{\lambda _{2j-1}} - {\lambda _{2j}}} \right|}.
\label{eq:delta_lambda}
\end{eqnarray}
Given sufficiently large $N_{\lambda}$, the condition $\delta
\lambda \to 0$ guarantees twofold degeneracy in the entanglement
spectrum, which is contributed by a pair of Majorana fermions
separated in each entanglement subsystem and hence regarded as a
signature for their existence (see details in
Refs.~\cite{Fidkowski10b,Turner11,Borchmann14}).

The second group involves thermodynamic quantities that can be
expressed as the derivative of energy density $E = \left\langle H
\right\rangle / L$ with respect to the system parameters. We focus
on the first derivative $-\frac{\partial E}{\partial \mu}$, which
is equal to the particle density $\rho  = N/L$ from the
Hellmann-Feynman theorem, and second derivatives $-
\frac{{{\partial ^2}E}}{{\partial X\partial Y}}$, which describe
compressibility ($X=Y=\mu$) or various susceptibility
($X=Y=\Delta$ or $X=\mu,Y=\Delta$) of the system. A discontinuity
or a peak of energy derivatives represents a drastic change in the
behavior of the ground-state energy, which implies a cross between
two lowest-energy states of the system. Such a cross is a
necessary condition for a topological transition.

The third group includes quantities describing condensate
properties. Superconducting fermions can be considered as a pair
condensate in which a two-body or Cooper-pair state is
macroscopically occupied~\cite{Yang62,Leggett06}, analogous to the
Bose-Einstein condensation in bosonic
systems~\cite{Penrose56,Sun09}. The condensate properties are well
characterized by the pair density matrix ${\rho ^{{\rm{pair}}}}$,
whose element is defined in spatial coordinates as
\begin{eqnarray}
\rho _{jk;j'k'}^{{\rm{pair}}} = \left\langle {c_j^\dag c_k^\dag
{c_{k'}}{c_{j'}}} \right\rangle.
\end{eqnarray}
When the condensation occurs, ${\rho ^{{\rm{pair}}}}$ has an
eigenvalue $\lambda _0^{{\rm{pair}}}$ [$\sim O(N)$] largely
compared to the others [$\sim O(1)$], representing the macroscopic
occupation or the number of condensed pairs. This number defines
the condensate fraction as
\begin{eqnarray}
P = \frac{{\lambda _0^{{\rm{pair}}} -
2}}{N},\label{eq:Condensate_fraction}
\end{eqnarray}
with an offset 2 of the noninteracting limit~\cite{Sun14a} [such
that $P=0$ for a free system described by $H_0$ in
Eq.~(\ref{eq:free_H})]. The eigenstate $\psi _{jk}^{{\rm{pair}}}$
corresponding to $\lambda _0^{{\rm{pair}}}$ represents the
Cooper-pair wave function. The Cooper-pair size is defined as a
root-mean-square distance between the two particles in a pair,
\begin{eqnarray}
{r_{{\rm{pair}}}} = \sqrt {\sum\limits_{j,k} {{{(j -
k)}^2}{{\left| {\psi _{jk}^{{\rm{pair}}}} \right|}^2}}
},\label{eq:Cooper-pair_size}
\end{eqnarray}
and indicates a length scale over which two fermions bind in real
space. The topological state is a weak-pairing state~\cite{Read00}
with ${r_{{\rm{pair}}}} \to \infty$ (or $\sim L$ in a finite-size
system), which originates the long-range entangled Majorana
fermions on both ends. Instead, the trivial state has finite
${r_{{\rm{pair}}}}$ (or $\ll L$). Therefore, a drastic change in
${r_{{\rm{pair}}}}$ is expectable upon the topological transition.
Note that the commonly used $U(1)$ symmetry-breaking
order~\cite{Alicea12} can distinguish the weak-pairing state
(being constant, corresponding to an infinite Cooper-pair size)
from the strong-pairing one (exponentially decaying in real space,
with the decay rate determining a finite Cooper-pair size) for a
number-nonconserving state, but fails to do so with a
number-conserving one (e.g., $\Delta \to 0$ in our model). The
definition of ${r_{{\rm{pair}}}}$ in
Eq.~(\ref{eq:Cooper-pair_size}) works for both. (In fact, such
${r_{{\rm{pair}}}}$ has been applied to study topological
properties in a two-dimensional system~\cite{Rombouts10}.)

We have shown the model Hamiltonian and all of the physical
quantities of interest. Among these quantities, twofold
degeneracies of ground-state energy and entanglement spectrum can
be regarded as direct signatures for the topological states and
Majorana fermions. They have also been proved for interacting
systems~\cite{Stoudenmire11,Turner11}. A large Cooper-pair size is
directly related to the long-range entanglement for the presence
of Majorana fermions, while the interaction effects on it are to
be investigated. The energy derivatives such as compressibility
and susceptibility can imply a cross of the two lowest-energy
states, but its relation to the topological phase transition needs
to be confirmed by the comparison with the direct signatures.
Below we will take the ground-state energy and entanglement
spectrum as a reference to pinpoint other signatures displayed in
the thermodynamic quantities and/or condensate properties. In the
next two sections, we present the numerical setup, results, and
discussions.

\vspace{-0.3cm}
\section{Numerical setup}\label{sec:numerics}
\vspace{-0.3cm}

While exact solutions can be found in specific parameter
regions~\cite{Katsura15}, the many-body ground state of the
interacting Hamiltonian in Eq.~(\ref{eq:Hamiltonian}) can, in
general, be computed numerically. Our numerical results are
obtained using the DMRG~\cite{White92,Schollwock05} method, whose
accuracy has been demonstrated in computing the ground-state
properties of a short-range coupled 1D system. This method has
been widely adopted to study topological properties in spinless
fermions~\cite{Thomale13,Iemini15}, spin-orbit coupled
electrons~\cite{Stoudenmire11,Chan15,Keselman15}, and ultracold
atoms~\cite{Kraus13,Liang14}. In our work, we employ the DMRG
method on systems up to $L=256$. The $Z_2$ symmetry of parity
conservation is considered to reduce the computational cost. We
keep up to $m=120$ states and apply seven sweeps in the
ground-state calculation. The number of states $m$ kept in the
DMRG method determines the size of the approximated Hamiltonian
and hence the accuracy of calculations. The discarded weight in
the eigenvalue of the reduced density matrix with $m=120$ is of
the order of $10^{-8}$, which guarantees the convergence of the
ground-state properties.

The DMRG calculations are numerically efficient for obtaining the
ground-state energy, entanglement spectrum, and real-space
two-point correlations. However, the bottleneck of our numerical
study lies in the computation of the condensate properties. The
construction of the pair density matrix requires the evaluation of
all possible four-point correlations. Due to the $L^4$-growing
computational cost as the system size increases, we are bounded to
$L=64$. Another limit is the diagonalization of the pair density
matrix, whose computational effort scales as $L^8$ and will
eventually dominate in a large-size case.

\vspace{-0.3cm}
\section{Results}\label{sec:results}
\vspace{-0.3cm}

In this section, we present and discuss results that track
multiple quantities for several cases of interest. We take the
tunneling strength $t$ to be the energy units for convenience and
set $\Delta = 0.2$. First, we study the well-known Kitaev chain
(where the internal interaction $V' = 0$) and find the
compressibility and susceptibility as good signatures for the
topological transition. Second, we track the compressibility for
general cases with the internal interaction. We will show that it
remains a valid indicator because it indicates the same
topological region as the entanglement spectrum does. Third, we
investigate the condensate properties together with the
compressibility as topological signatures for a relatively small
system. We also study the characterization of topological phases
with the use of number density and condensed-pair density.
Finally, we present the effects of finite size, disorder, and
inhomogeneity.

\vspace{-0.3cm}
\subsection{The Kitaev model}
\vspace{-0.3cm}

The Kitaev chain model is described by the Hamiltonian of
Eq.~(\ref{eq:Kitaev}). It also represents the weakly interacting
limit $V \to 0$ of our model Hamiltonian of
Eq.~(\ref{eq:Hamiltonian}). The upper and lower topological
transition points of an infinite Kitaev chain are known as $\mu =
\pm 2 \equiv \mu_c^\pm$~\cite{Kitaev01}, between (outside) which
the system is in a topological (trivial) state. For a finite
chain, we numerically calculate the transition points. They show
little change if the system size is large compared with the
characteristic size of Majorana fermions.

\begin{figure}[t]
\centering
\includegraphics[width=8.6cm]{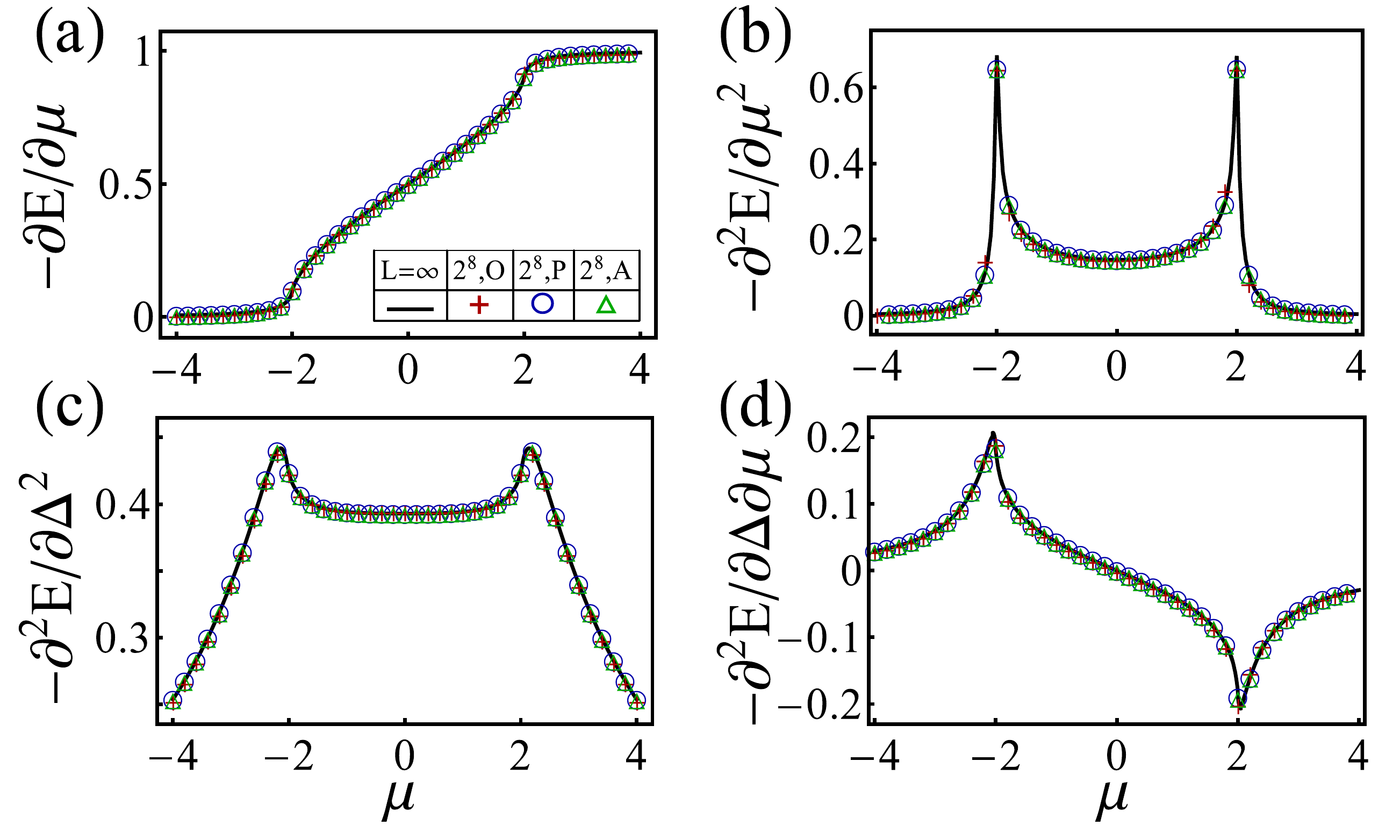}\vspace{-0.3cm}
\caption{(Color online) Noninteracting system energy's (a) first
derivative $-\frac{\partial E}{\partial \mu}=\rho$ and various
second derivatives (b) $-\frac{\partial^2 E}{\partial \mu^2}=
\frac{\partial \rho}{\partial \mu}$, (c) $-\frac{\partial^2
E}{\partial \Delta^2}$, and (d) $-\frac{\partial^2 E}{\partial \mu
\partial \Delta}=\frac{\partial \rho}{\partial \Delta}$, vs chemical potential $\mu$. Four cases of $L=\infty$
(solid curve) and $L=256$ with open (red crosses), periodic (blue
circles), and antiperiodic (green triangles) boundary conditions
are presented. The superconducting gap is set as $\Delta = 0.2$.}
\label{fig:f1}
\end{figure}

Figure \ref{fig:f1} shows four thermodynamic quantities as energy
derivatives: density $\rho=-\frac{\partial E}{\partial \mu}$
[Fig.~\ref{fig:f1}(a)], compressibility $\frac{\partial
\rho}{\partial \mu}=-\frac{\partial^2 E}{\partial \mu^2}$
[Fig.~\ref{fig:f1}(b)], and susceptibilities $-\frac{\partial^2
E}{\partial \Delta^2}$ [Fig.~\ref{fig:f1}(c)] as well as
$-\frac{\partial^2 E}{\partial \Delta
\partial \mu}$ [Fig.~\ref{fig:f1}(d)], vs the chemical potential $\mu$. We plot data for
an infinite chain (solid curve) and a finite chain of $L=256$ with
open (red crosses), periodic (blue circles), and antiperiodic
(green triangles) boundary conditions. We see that the density
smoothly varies from vacuum ($\rho=0$) to commensurate filling
($\rho=1$) as $\mu$ increases. The trivial states have either a
very low filling ($\rho \sim 0$ in $\mu < \mu_c^-$) or an almost
commensurate filling ($\rho \sim 1$ in $\mu > \mu_c^+$). All of
the second derivatives of the energy develop peaks at each
transition point. The compressibility peaks reflect the bulk gap
closing at the transition point~\cite{Nozadze15}. Across the peak,
a sign change of the curve slope indicates a discontinuity in the
third derivative of energy, which implies that the topological
transition is third order. Such a topological transition type has
also been found in a long-range coupled system~\cite{Ortiz14}. The
peak structure is independent of the chain size (see more analyses
in Sec.~\ref{sec:results} D) as well as the boundary conditions
and can thus be taken as a reliable signature for the topological
transition. While the two susceptibilities both ought to work
well, below we study the compressibility, i.e., a typical
observable in experiments, as an indicator for cases with internal
interaction.

\vspace{-0.3cm}
\subsection{Effects of internal interaction}
\vspace{-0.3cm}

\begin{figure*}[t]
\centering
\includegraphics[width=18cm]{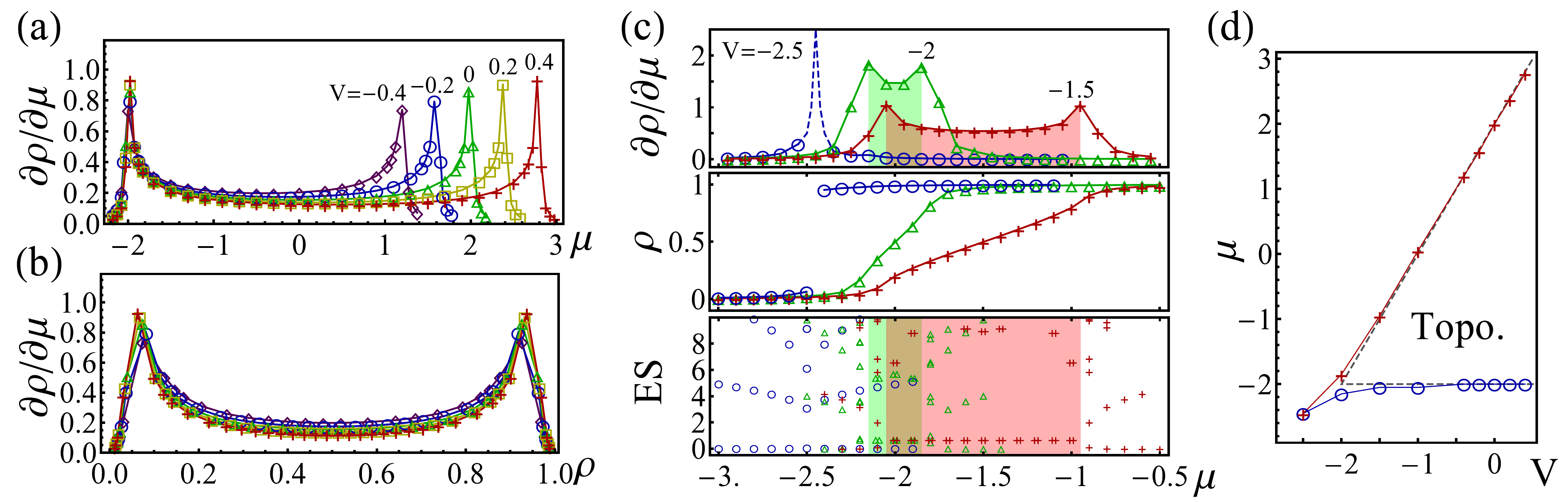}\vspace{-0.3cm}
\caption{(Color online) (a),(b) Compressibility $\frac{\partial
\rho}{\partial \mu}$ vs $\mu$ and $\rho$, respectively, at various
interactions $V=-0.4$ (purple diamonds), $-0.2$ (blue circles),
$0$ (green triangles), $0.2$ (yellow squares), and $0.4$ (red
crosses) for an open chain of $L=256$ and $\Delta=0.2$. (c)
$\frac{\partial \rho}{\partial \mu}$ (top panel), $\rho$ (middle
panel), and entanglement spectrum (ES, bottom panel) vs $\mu$ at
relatively large attraction $V=-1.5$ (red crosses), $-2$ (green
triangles), and $-2.5$ (blue circles). The topological region for
each $V$ is shaded in the top and bottom panels for comparison.
The dashed curve in the top panel shows how the two
compressibility peaks would emerge at $V=-2.5$, although the
actual value is undefined due to the discontinuity in $\rho$. (d)
Interacting phase diagram showing the topological region in the
$V$-$\mu$ plane. The upper and lower boundaries are marked by red
crosses and blue circles, respectively, while the dashed curve
shows the Hartree mean-field results.} \label{fig:f2}
\end{figure*}

With the internal interaction turned on ($V \neq 0$), we calculate
the ground state properties as a function of $V$ for a case of
$L=256$. In Fig.~\ref{fig:f2}(a), we plot compressibility
$\frac{\partial \rho}{\partial \mu}$ vs chemical potential $\mu$
at various $V=-0.4$ (purple diamonds), $-0.2$ (blue circles), $0$
(green triangles), $0.2$ (yellow squares), and $0.4$ (red
crosses). The curves maintain the two-peak structure, while the
right transition point $\mu_c^+$ has a positive (negative) shift
at repulsive (attractive) internal interaction, and the left one
$\mu_c^-$ shows little change with the interaction. We also
confirm that the region sandwiched by the compressibility peaks
coincides with the topological region identified by the
ground-state degeneracy and entanglement spectrum. Therefore,
compressibility can be regarded as a reliable signature for
topological states of interacting systems. Figure ~\ref{fig:f2}(b)
shows compressibility vs density $\rho$ in the same convention as
Fig.~\ref{fig:f2}(a). Similarly, the curves display peaks at the
two transition points. We notice that the position of the left
(right) peak is at a low (high) filling, $\rho < 0.1$ ($>0.9$),
and is insensitive to interaction. In other words, the topological
state survives in most intermediate-filling regions, which
promises the Majorana fermions in a wide range of
density-controllable systems such as ultracold atoms.

Since the topological region shrinks as the interaction becomes
more attractive, we turn to study the fate of the topological
state in the strongly attractive region. (Note that the strongly
repulsive region has been well studied in Ref.~\cite{Thomale13}.)
The top panel of Fig. \ref{fig:f2}(c) shows compressibility vs
$\mu$ at $V=-1.5$ (red crosses), $-2$ (green triangles), and
$-2.5$ (blue circles), with the topological region shaded. We see
that the two peaks at $\mu_c^\pm$ merge into one at $V=-2.5$ and
the topological region disappears. The disappearance can also be
seen in the entanglement spectrum in the bottom panel as the
doubly degenerate region vanishes at $V=-2.5$. The density curves
in the middle panel show that the topological transitions at
$V=-1.5$ and $-2$ still happen at either a low or high filling.
When the topological region disappears at $V=-2.5$, the density
($=-\frac{\partial E}{\partial \mu}$) curve exhibits a
discontinuity and the system undergoes a first-order transition
between a trivial low-filling state and a trivial high-filling
one.

The phase boundary shift can be explained by an effective
chemical-potential shift due to the interaction. Considering a
Hartree approximation ${n_{j + 1}}{n_j} \to \rho {n_j} +\rho
{n_{j+1}} -\rho^2$, one can turn the Hamiltonian of
Eq.~(\ref{eq:Hamiltonian}) into the original Kitaev form with an
effective chemical potential ${\mu _{\rm{eff}}} = \mu - 2V\rho$,
where $\rho=\rho(\mu,V )$. The upper and lower transition points
are thus given by ${\mu _{\rm{eff}}}=\pm 2$, respectively, which
leads to
\begin{eqnarray}
\mu _c^ \pm  =  \pm 2 + 2V\rho (\mu _c^ \pm,V
).\label{eq:phase_boundary}
\end{eqnarray}
Because the density at the lower (upper) transition point is low
(high) and insensitive to interaction, we approximately insert
$\rho (\mu _c^-,V )=0$ and $\rho (\mu _c^+,V )=1$ into
Eq.~(\ref{eq:phase_boundary}) and obtain $\mu _c^ + = 2+2V$ and
$\mu _c^ - = -2$. Such relations tell that the lower boundary
barely depends on $V$ and the upper boundary linearly shifts with
$V$. Figure \ref{fig:f2}(c) shows the topological region in the
$V$-$\mu$ plane. The upper and lower phase boundaries from this
approximation (dashed curves) well match those from the numerical
calculations (red crosses and blue circles, respectively).

\vspace{-0.3cm}
\subsection{Condensate properties}
\vspace{-0.3cm}

\begin{figure}[b]
\centering
\includegraphics[width=7.5cm]{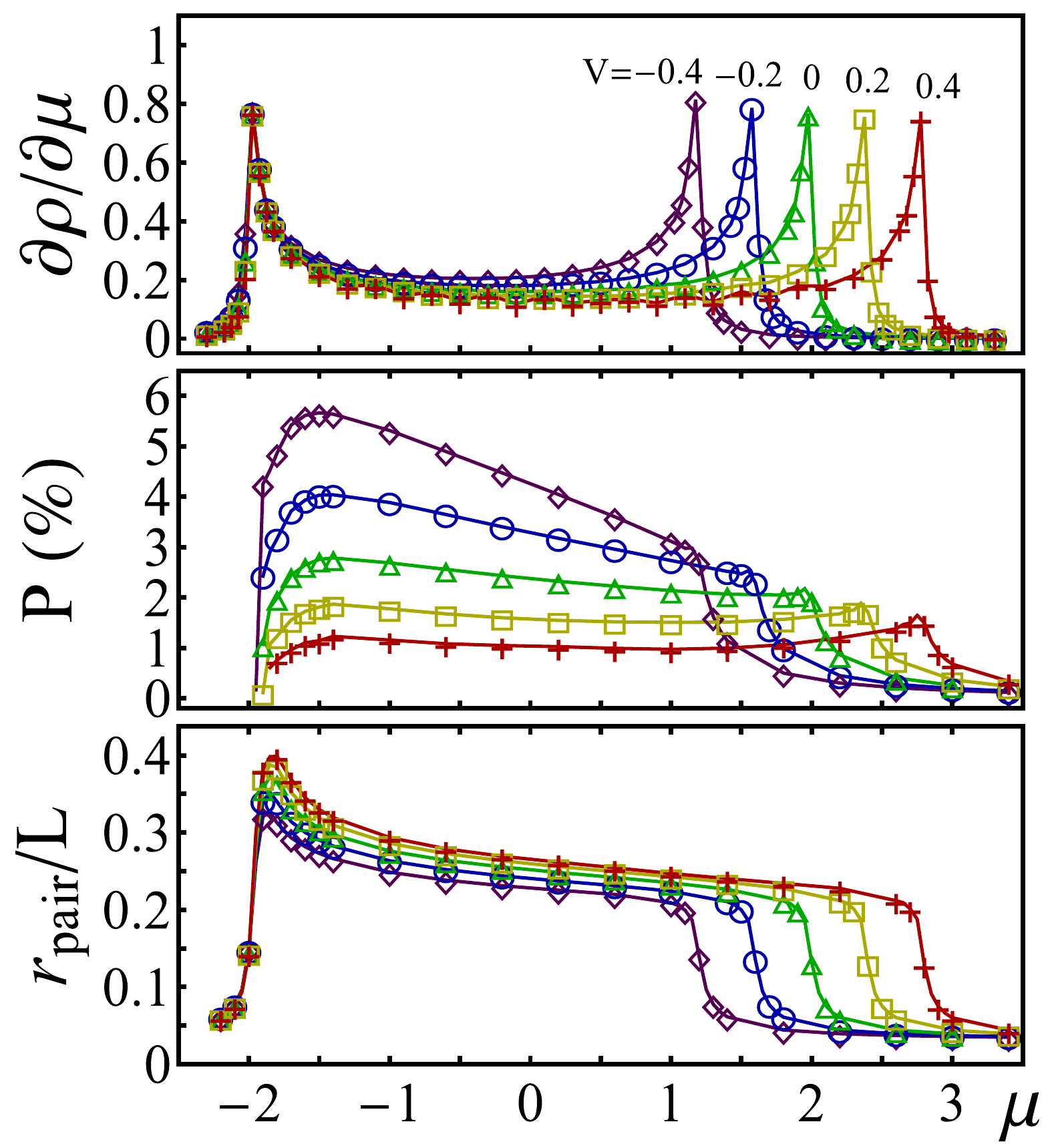}\vspace{-0.3cm}
\caption{(Color online) Compressibility $\frac{\partial
\rho}{\partial \mu}$ (top panel), condensate fraction $P$ (middle
panel), and Cooper-pair size $r_{\rm{pair}}$ (bottom panel) vs
$\mu$ at various interactions for an open chain of $L=32$ and
$\Delta = 0.2$. Conventions are the same as in
Fig.~\ref{fig:f2}(a).} \label{fig:f3}
\end{figure}

In this section, we study pair-condensate properties of the
system. As mentioned in Sec.~\ref{sec:numerics}, the calculation
of the pair density matrix can be time consuming. Such a
constraint directs our focus on a relatively small system ($L=32$)
rather than a large one. The top panel of Fig.~\ref{fig:f3} shows
the compressibility curves as a reference for the topological
region at various interactions [conventions are the same as in
Fig.~\ref{fig:f2}(a)]. The middle and bottom panels show the
condensate fraction $P$ defined in
Eq.~(\ref{eq:Condensate_fraction}) and Cooper-pair size
$r_{\rm{pair}}$ defined in Eq.~(\ref{eq:Cooper-pair_size}),
respectively. Different from the compressibility and entanglement
spectrum, we first see that the curves of condensate properties
lack symmetry with respect to the half-filling point. At the upper
phase boundary (right peak of the compressibility curve) both $P$
and $r_{\rm{pair}}$ develop a kink, indicating a sudden change in
the condensate properties upon the topological transition. The
sharp decrease of $r_{\rm{pair}}$ is consistent with the
transition from weak-pairing (topological) to strong-pairing
(trivial) states~\cite{Read00,Alicea12}. At the lower boundary
(left peak of the compressibility curve), $r_{\rm{pair}}$ shows a
peak, while $P$ changes the trend but does not show a clear
signature due to the finite-size effect (the filling is so low
such that the total number of particles is smaller than two). We
increase the system size and find that the turning point of $P$
approaches the phase boundary. Therefore, the behavior of $P$ and
$r_{\rm{pair}}$ can be an indicator for the topological
transition.

The topological transition accompanied by a significant change in
the Cooper-pair size raises an interesting question of whether the
change in the pair condensate can be purely qualitative or must be
both qualitative and quantitative. We try to answer this question
by examining two essential quantities of a pair condensate, i.e.,
the total number of particles $N$ and the total number of
condensed pairs $\lambda_0^{\rm{pair}}$, around the topological
phase boundary. The answer is the former if the system can go
across the phase boundary and keep both $N$ and
$\lambda_0^{\rm{pair}}$ unchanged (only $r_{\rm{pair}}$ changes).
Our model is suited to explore this question because it is defined
in a three-dimensional parameter space $(\Delta,V,\mu)$ such that
a path along which two functions $N(\Delta,V,\mu)$ and
$\lambda_0^{\rm{pair}}(\Delta,V,\mu)$ remain constant becomes
mathematically possible. We explore large enough regions around
upper and lower transition points, $(\Delta,V,\mu)=(0.2,0,\pm 2)$,
respectively, by varying all three parameters for a $L=64$ case.
The results show that one of $N$ and $\lambda_0^{\rm{pair}}$ can
remain constant upon the topological transition, but not both. In
other words, the topological transition or the sudden change of
the Cooper-pair size in our model can be regarded as a result of a
quantitative change in $N$ or $\lambda_0^{\rm{pair}}$. Moreover,
we do not find that a topological state and a trivial state have
the same $N$ and $\lambda_0^{\rm{pair}}$. Our results may have
implications for the characterization of topological states by
$(N/L,\lambda_0^{\rm{pair}}/L)$.

\begin{figure}[t]
\centering
\includegraphics[width=7.5cm]{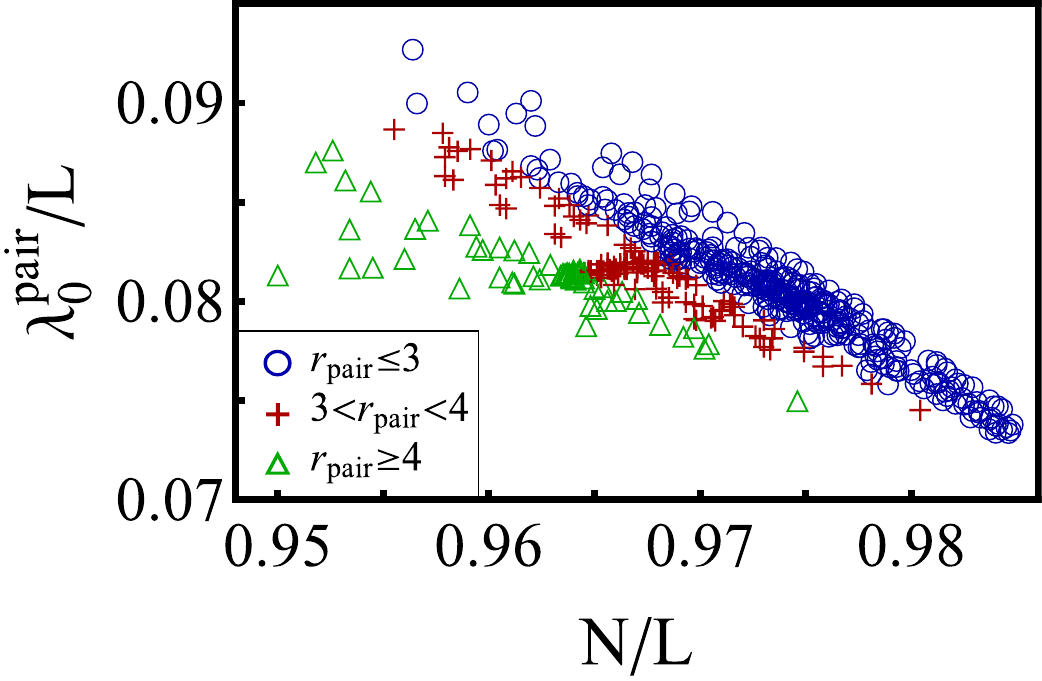}\vspace{-0.3cm}
\caption{(Color online) Data points showing three different ranges
of the Cooper-pair size, i.e., $r_{\rm{pair}} \le 3$ (blue
circles), $3 < r_{\rm{pair}} <4 $ (red crosses), and
$r_{\rm{pair}} \ge 4$ (green triangles), in the plane of $N/L$ and
$\lambda_0^{\rm{pair}}/L$ for $L=32$.} \label{fig:f4}
\end{figure}

We further study this possibility by computing the condensate
properties for hundreds of points in the range of $0.15 \le \Delta
\le 0.45$, $-0.45 \le V \le 0.4$, and $1.4 \le \mu \le 3.4$, for
$L=32$. Figure \ref{fig:f4} shows the data points representing
three different ranges of the Cooper-pair size around the upper
topological transition region in the plane of $N/L$ and
$\lambda_0^{\rm{pair}}/L$. The system can be considered as a
topological state if $r_{\rm{pair}} \ge 4$. We see that along a
vertical (horizontal) path in Fig.~\ref{fig:f4}, the system can
undergo a topological transition at fixed $N$
($\lambda_0^{\rm{pair}}$). On the other hand, the concurrence of a
decrease in $r_{\rm{pair}}$ and an increase in either $N$ or
$\lambda_0^{\rm{pair}}$ confirms our conjecture, i.e., an
inevitable quantitative change upon the topological transition. In
other words, $r_{\rm{pair}}$ is a function of $N$ and
$\lambda_0^{\rm{pair}}$, and these two parameters can hence be
used to characterize the topological phase diagram. We finally
comment that a more solid confirmation lies in the convergence of
Fig.~\ref{fig:f4} in the thermodynamic limit; we leave the test of
this for future study once more powerful computational tools are
available.

\vspace{-0.3cm}
\subsection{Effects of finite size, disorder, and inhomogeneity}
\vspace{-0.3cm}

\begin{figure}[t]
\centering
\includegraphics[width=7.5cm]{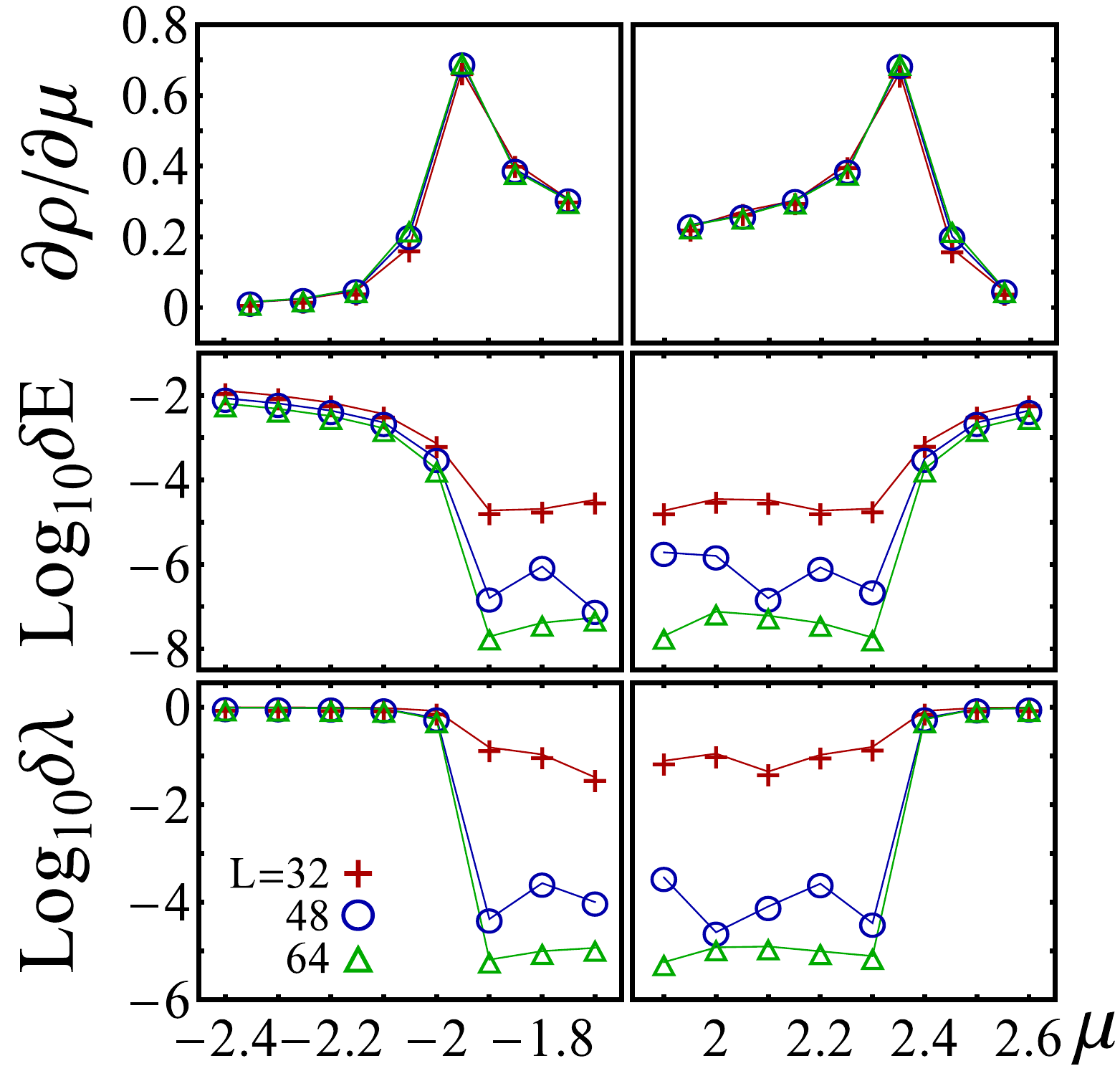}\vspace{-0.3cm}
\caption{(Color online) Compressibility $\frac{\partial
\rho}{\partial \mu}$ (top panel), energy gap between the ground
state and first excited state $\delta E$ (middle panel), and
difference between even and odd sectors of entanglement spectrum
$\delta \lambda$ (bottom panel) vs $\mu$ around the lower (left
column) and upper (right column) transition points at various
open-chain sizes: $L=32$ (red crosses), $48$ (blue circles), and
$64$ (green triangles), respectively. Data are for $\Delta = V =
0.2$.}\vspace{-0.3cm} \label{fig:f5}
\end{figure}

In the last section, we explore the stability of the signatures
against three realistic effects in experiments---finite size,
disorder, and inhomogeneity. Figure \ref{fig:f5} shows the
compressibility (top panel), energy difference $\delta E$ (middle
panel) defined in Eq.~(\ref{eq:delta_E}), and entanglement
spectrum difference $\delta \lambda$ (bottom panel) defined in
Eq.~(\ref{eq:delta_lambda}) vs $\mu$ for $L=32$ (red crosses),
$48$ (blue circles), and $64$ (green triangles). We see that
although all three quantities indicate the transition points, the
compressibility curve is insensitive to the system size, while
$\delta E$ and $\delta \lambda$ can increase by several orders as
the system size is halved. Therefore, the compressibility is a
stable indicator against the finite-size effect. The insensitivity
also provides a more numerically efficient way to predict the
topological transition of a large-size system by calculating the
compressibility peak of a relatively small one.

\begin{figure}[t]
\centering
\includegraphics[width=7.5cm]{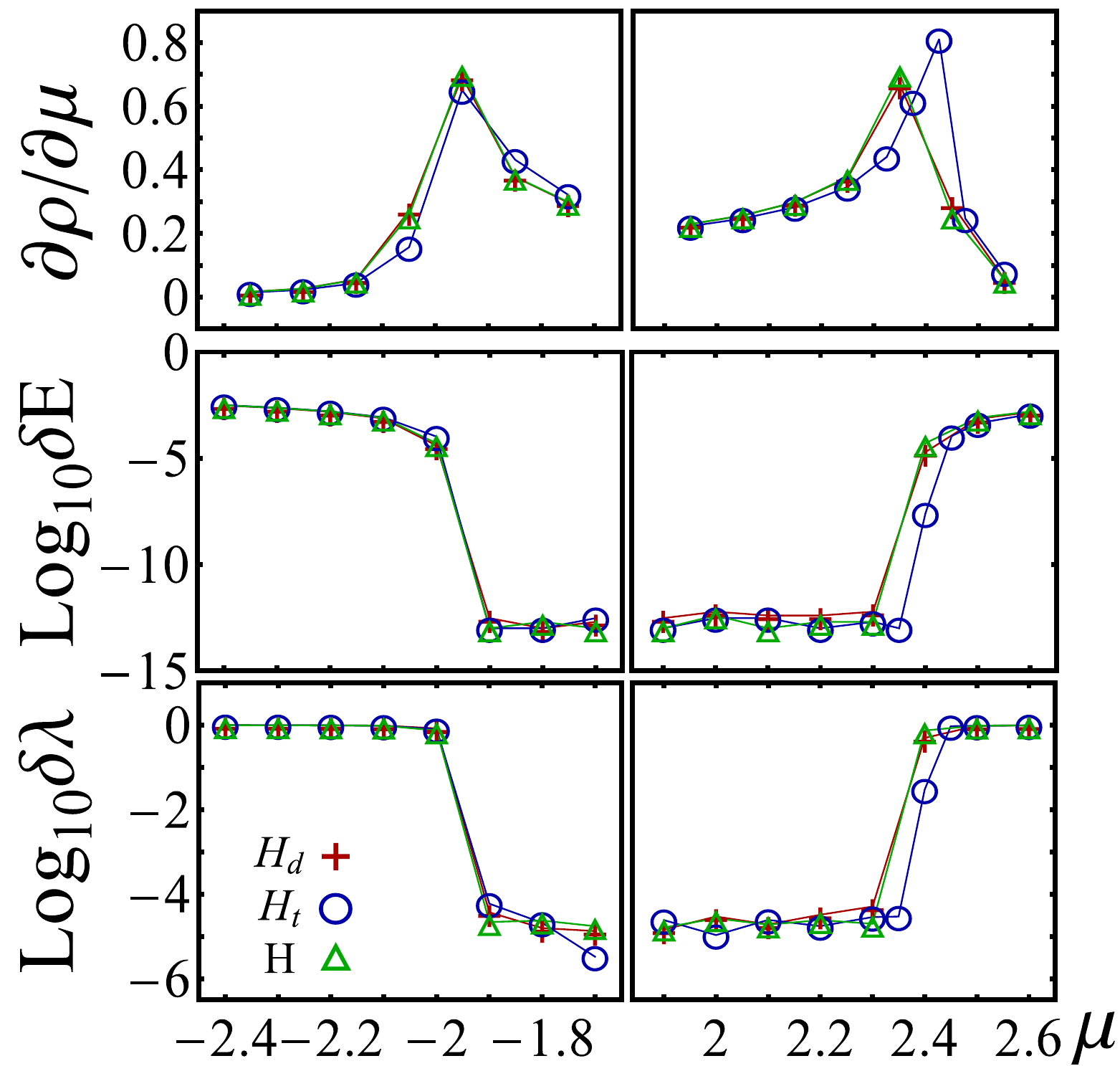}\vspace{-0.3cm}
\caption{(Color online) Comparison between a disorder system
[$H_d$ of Eq.~(\ref{eq:disorder}), red crosses], a trap system
[$H_t$ of Eq.~(\ref{eq:trap}), blue circles], and a reference
without these effects [$H$ of Eq.~(\ref{eq:Hamiltonian}), green
triangles]. Conventions are the same as in Fig.~\ref{fig:f5},
except the system size here is $L=128$.} \label{fig:f6}
\end{figure}

For the disorder effect, we consider a Hamiltonian
\begin{eqnarray}
{H_d} = H + \sum\limits_{j = 1}^L {{\delta _j}{{\hat n}_j}},
\label{eq:disorder}
\end{eqnarray}
with a set of random local potential shifts $\{ \delta_j \}$ that
obey the normal distribution and have zero average. For the
inhomogeneous effect, we consider an external harmonic trap with
curvature $K$ turned on,
\begin{eqnarray}
{H_t} = H + \sum\limits_{j = 1}^L {\frac{K}{2}{{\left( {j -
\frac{L}{2} - \frac{1}{2}} \right)}^2}{{\hat
n}_j}}.\label{eq:trap}
\end{eqnarray}
In Fig.~\ref{fig:f6}, we compare a disorder case (red crosses)
with the variance of $\{ \delta_j \}$ equal to $0.1$ and a trap
case (blue circles) with $K=0.8/(L-1)^2$ with the original
Hamiltonian $H$ for $L=128$ in the same convention of
Fig.~\ref{fig:f5}. We see that neither of the two effects can
alter the signatures for the topological transition. Such results
are anticipated because the topological states and Majorana
fermions are symmetry protected. Perturbations can not destroy the
topological order as long as they are not strong enough to cause
the bulk-gap crossing. (Regimes of strong disorder or deep traps
are beyond the scope of this study. One could refer to previous
works in
Refs.~\cite{Ruhman15,Brouwer11,Lobos12,DeGottardi13,Crein14}.)

\vspace{-0.5cm}
\section{Conclusion}\label{sec:conclusion}
\vspace{-0.5cm}

In conclusion, multiple physical quantities have been analyzed for
one-component fermions with proximity-induced superconducting gap
and interparticle interaction in 1D lattices, which can be a
topological superconductor hosting Majorana fermions. In addition
to the double degeneracy of ground-state energy and entanglement
spectrum, we have found that the topological transition can also
be revealed by peaks of compressibility and susceptibility curves,
as well as a sudden change of trend in the condensate fraction and
Cooper-pair size. Among them, the compressibility peak is
particularly useful for its stability against the finite-size
effect and being observable in experiments. The Cooper-pair size
directly shows the topological transition between strong-pairing
and weak-pairing states. By tracking these signatures, we have
found that the topological transition is third order. As the
interaction becomes more attractive, the topological state finally
disappears and the system undergoes a first-order transition
between low-filling and high-filling trivial states. We have also
explored the possibility to characterize the topological phase
using density of particles and that of condensed pairs. One future
direction is the extension of this study to other interacting
platforms in which various tunneling or pairing
channels~\cite{Sun14a,Burset14,Liu15} need to be considered and an
alternative treatment suited for the continuous
space~\cite{Fidkowski11,Chung15} may apply. In addition, our
results may find applications on spin systems that are associated
with our Hamiltonian of Eq.~(\ref{eq:Hamiltonian}), such as an
Ising chain with a transverse field ($V \to 0$, e.g., see
Ref.~\cite{Perk77}), the $XXZ$ model ($\Delta \to
0$)~\cite{Yang66}, the Baxter $XYZ$ model ($\mu \to
0$)~\cite{Baxter82}, and the two-dimensional classical Ising model
($V \to 0$, $t=\Delta$)~\cite{Kaufman49}.

\textbf{Acknowledgements}: We are grateful to J.~Alicea,
C.~J.~Bolech, L.~Duan, M.~Franz, T.~L.~Hughes, H.-H.~Hung,
A.~J.~Leggett, J.~H.~H.~Perk, N.~Shah, and C.~Zhang for
informative discussions. Y.H.C.~acknowledges the support by the
Thematic Project at Academia Sinica. C.K.C.~gratefully
acknowledges the support of the Max-Planck-UBC Center for Quantum
Materials. K.S.~is supported by ARO (Grant No.~W911NF-12-1-0334)
and AFOSR (Grant No.~FA9550-13-1-0045). Part of the numerical work
was developed using the DMRG code released within the Powder with
Power project (http://qti.sns.it/dmrg/). We acknowledge the
computational resource provided by University of Cincinnati and
Texas Advanced Computing Center (TACC).

\end{document}